# Spin and valley polarized one-way Klein tunneling in photonic topological insulators

Xiang Ni[1,2*], David Purtseladze[3*], Daria A. Smirnova[1*], Alexey Slobozhanyuk[1], Andrea Alù[3], and Alexander B. Khanikaev[1,2]

[1]Department of Electrical Engineering, Grove School of Engineering, The City College of the City University of New York, 140th Street and Convent Avenue, New York, NY 10031, USA

[2]Graduate Center of the City University of New York, New York, NY 10016, USA

[3]Department of Electrical and Computer Engineering, The University of Texas at Austin, Austin, TX 78701, USA

E-mail: khanikaev@gmail.com

*These authors contributed equally to the present work

**Advances of condensed matter physics in exploiting the spin degree of freedom of electrons led to the emergence of the field of spintronics, which envisions new and more efficient approaches to data transfer, computing, and storage [1-3]. These ideas have been inspiring analogous approaches in photonics, where the manipulation of an artificially engineered pseudo-spin degrees of freedom is enabled by synthetic gauge fields acting on light [4,5,6]. The ability to control these additional degrees of freedom can significantly expand the landscape of available optical responses, which may revolutionize optical computing and the basic means of controlling light in photonic devices across the entire electromagnetic spectrum. Here we demonstrate a new class of photonic systems, described by effective Hamiltonians in which competing synthetic gauge fields engineered in pseudo-spin, chirality/sublattice and valley subspaces result in band gap opening at one of the valleys, while the other valley exhibits Dirac-like conical dispersion. It is shown that such effective response has dramatic implications on photon transport, among which: (i) spin-polarized and valley-polarized one-way Klein tunneling and (ii) topological edge states that coexist within the Dirac continuum for opposite valley and spin polarizations. These phenomena offer new ways to control light in photonics, in particular for on-chip optical isolation, filtering and wave-division multiplexing by selective action on their pseudo-spin and valley degrees of freedom.**

The rich physics of electromagnetic phenomena facilitated by the internal degrees of freedom of light, e.g., polarization and angular momentum, has enabled a multitude of applications from multimode optical fibers and holography, to tractor beams and topologically robust propagation [7-9]. This plethora of opportunities for optical manipulation is enabled by the use of synthetic photonic potentials implemented via engineered light-matter interactions in structured materials, which allow imprinting specific structure to light. Periodically patterned optical media, such as photonic crystals and metamaterials [10,11], play an important role here, by allowing the emulation of a wide range of condensed matter phenomena. Indeed, suitably tailored periodicity has enabled the emergence of photonic bandgaps and most recently the discovery of photonic topological insulators [4-6,12-27], enabling parallels with condensed matter physics that have been an inspirational driving force for photonics research.

Recent progress in spintronics research has opened new opportunities in manipulating spin-polarized carriers, being one additional example of such inspiration. Indeed, in photonics engineered pseudo-spin has been of significant interest as an additional degree of freedom in the context of topological electromagnetic states [4-6,12-27]. Photonic structures with synthetic gauge potentials selectively acting on the polarization of light, its angular momentum or valley degrees of freedom have been used to engineer symmetry protected topological order for photons and topologically robust pseudo-spin polarized transport [24-27]. More recently, coupling of spin and valley degrees of freedom has been employed to create spin-polarized bandgaps at the two valleys of a photonic analogue of graphene [27].

To engage the full potential of spin-valley photonics, we should design a special class of materials exhibiting disparate metallic-like vs insulating states for opposite photonic pseudo-spins and valleys. In this manuscript, we achieve such selectivity with respect to both spin and valley degrees of freedom and non-trivially extend the space of available effective photonic potentials by coupling spin, valley and sublattice (chiral) degrees of freedom. We consider systems with and without time-reversal symmetry (TRS), demonstrating the existence of one-way valley-polarized and spin-valley polarized Dirac spectra. The proposed structures effectively behave as Dirac semimetals and insulators for opposite pseudo-spin and/or valley polarizations, respectively. Such exotic responses emerge at the crossover between distinct topological photonic phases: i) quantum Hall and valley-Hall, ii) quantum spin-Hall and valley-Hall phases. The critical points of topological transitions are found to exhibit unusual properties at topological domain walls, where the pseudo-spin and valley polarized edge states coexist with the continuum associated with the opposite polarizations.

The systems we consider here, shown in Fig. 1a and Fig. 1b, emulate two different scenarios without and with TRS and represent photonic crystals with hexagonal and triangular lattices, respectively. The degeneracies with crossed linear Dirac-like dispersions in honeycomb crystals are protected by both spatial inversion symmetry (SIS) and TRS. In photonic lattices, such Dirac cones necessarily appear in pairs, unless TRS is broken. For this reason, photonic graphene exhibits two pairs of Dirac cones at two valleys of its Brillouin zone [28]. Provided that the intervalley scattering is absent, the valleys can be used as an additional degree of freedom, offering a new approach to engineer and control electrons and photons [29-48]. In particular, coupling of valley and spin degrees of freedom has been proposed to generate spin-valley-coupled currents and for spin-valley filtering in condensed matter physics. In photonics, the valley degree of freedom has been used to emulate spin in photonic topological insulators [6], and for separation of pseudo-spin flows in metamaterials [27].

**I. Topological photonic crystals with one-way Dirac spectra**

We first consider the simpler case of photonic graphene (Fig.1a) with broken TR symmetry – a honeycomb array of high-index ferromagnetic rods magnetized along their axes. The bipartite unit cell consists of cylinders with radii $r_A$ and $r_B$ with scalar permittivity $\epsilon$, in a lattice with lattice constant $a$. The gyromagnetic response of the ferromagnetic material is induced by a static magnetic field $B_0$ along the z-axis and it is described by a magnetic permeability tensor with diagonal components $\mu_{xx} = \mu_{yy} = \mu$ and off-diagonal components $\mu_{xy} = -\mu_{yx} = i\kappa$. Two types of symmetry reduction can be introduced to this system, resulting in two distinct quantum-Hall-like [12,13,15] and valley-Hall topological phases [12,47]. In the photonic graphene, the valley-Hall phase is induced by the reduction of the sublattice symmetry due to the dimerization of A and B sites, i.e., by making the radii unequal. The quantum-Hall-like phase is attained by breaking of TR symmetry due to magnetization. For the photonic crystals with hexagonal symmetry considered here, the spectrum hosts a pair of Dirac points at K and K′ valleys of the Brillouin zone. Both symmetry reductions induce synthetic gauge fields leading to the inversion of bands touching at the Dirac points. As a result, the bands acquire non-vanishing Berry curvature and can be ascribed an appropriate topological invariant – Chern or valley-Chern number for magnetized and dimerized lattices, respectively [12,31].

The effective Hamiltonian describing both states can be obtained from Maxwell's equations using the plane wave expansion method [11,4] and the k.p approximation in the vicinity of the Dirac points. Including the valley degree of freedom, the effective Hamiltonian then assumes the form (Supplement I, Section A)

$$\widehat{\mathcal{H}} = v_D \hat{\sigma}_x \hat{\tau}_z k_x + v_D \hat{\sigma}_y \hat{\tau}_0 k_y + \hat{\sigma}_z (\hat{\tau}_z m_T - \hat{\tau}_0 m_I), \qquad (1)$$

where $\hat{\tau}_i$ and $\hat{\sigma}_i$ are Pauli matrices in valley and sub-lattice degrees of freedom, and $m_T$ and $m_I$ are mass terms induced by TRS and SIS reductions, respectively.

The form of Hamiltonian (1) reveals that the effective mass due to TRS breaking ($m_T$-term) has opposite signs at the two valleys. On the other hand, the effective mass due to SIS breaking ($m_I$-term) has the same sign at both valleys. This implies that one can artificially tune the structure parameters to make the masses equal $m_T = m_I$, thus closing the bandgap at the K point and at the same time doubling it at the K′ point. As a result, the dispersion at the K′ valley exhibits locally quadratic (parabolic) dispersion $\Omega_\pm(\mathbf{k}) = \pm\sqrt{v_D^2 (k_x^2 + k_y^2) + m^2}$, with $m = m_T + m_I$, while the dispersion at the K valley exhibits gapless linear Dirac-like dispersion.

The second model we consider implements TRS preserving scenario with a pseudo-spin degree of freedom. It represents a triangular array of triangulated bi-anisotropic rods. The pseudo-spin degree of freedom is introduced by the duality symmetry, ensured by equal electric permittivity $\hat{\epsilon}$ and permeability $\hat{\mu}$ of the rods [5]. Such dual crystals have been shown to emulate quantum spin Hall state for the pseudo-spins $\psi^{\uparrow(\downarrow)} = E_z \pm H_z$ when the bianisotropic response is introduced. The bianisotropy is described by the effective constitutive relations with magneto-electric coupling $\mathbf{D} = \hat{\epsilon}\mathbf{E} + \hat{\xi}\mathbf{H}$ and $\mathbf{B} = \hat{\mu}\mathbf{H} + \hat{\xi}^\dagger\mathbf{E}$, where the only non-vanishing elements of the bianisotropy parameter $\hat{\xi}$ are $\xi_{xy} = -\xi_{yx} = i\Delta$.

The duality of the crystal provides double degeneracy of the spectrum with respect to TE and TM modes, while its triangular symmetry ensures the presence of two overlaid Dirac points for dipolar (orbital number $l = \pm 1$) bands at each valley. The bianisotropic response has an effect analogous to spin-orbit coupling in electronic systems, and it results in band-crossing of TE and TM bands and opening of topological band gaps for both pseudo-spins and valleys [5]. In addition, the Dirac points can be gapped by the reduction of spatial symmetry by triangulation of the rods, which has an effect analogous to the dimerization of a honeycomb (graphene) lattice giving rise to valley-Hall photonic state [24,47-48].

The effective Hamiltonian describing this structure can be again obtained from Maxwell's equations (see Supplement I, Section B) and has the form

$$\hat{\mathcal{H}} = v_\mathrm{D}\hat{\sigma}_x\hat{\tau}_z\hat{s}_0 k_x + v_\mathrm{D}\hat{\sigma}_y\hat{\tau}_0\hat{s}_0 k_y + \hat{\sigma}_z(\hat{\tau}_z\hat{s}_z m_\mathrm{B} - \hat{\tau}_0\hat{s}_0 m_\mathrm{I}), \qquad (2)$$

where an additional set of Pauli matrices $\hat{s}_i$ corresponding to the spin degree of freedom is introduced, and $m_\mathrm{B}$ and $m_\mathrm{I}$ are mass terms induced by the bianisotropy and SIS reductions, respectively.

The form of Hamiltonian (2) suggests that the two mass terms exhibit different behaviors at the two valleys, implying that one can again artificially tune the structure parameters to equate the effective masses $m_\mathrm{B} = m_\mathrm{I}$, thus closing (doubling) the band gap at K point, while doubling (closing) it at K' point for spin-up (spin-down) state. Thus, for this system TRS invariance ensures that opposite spins exhibit Dirac-like dispersion at the two opposite valleys. As a result, the dispersion for the spin-up (spin-down) state is quadratic at the K' (K) valley, with effective mass $m = m_\mathrm{B} + m_\mathrm{I}$. Both proposed systems therefore exhibit valley-selective linear Dirac-like dispersions, which should manifest itself in peculiar wave-transport properties both in the bulk and on the edges.

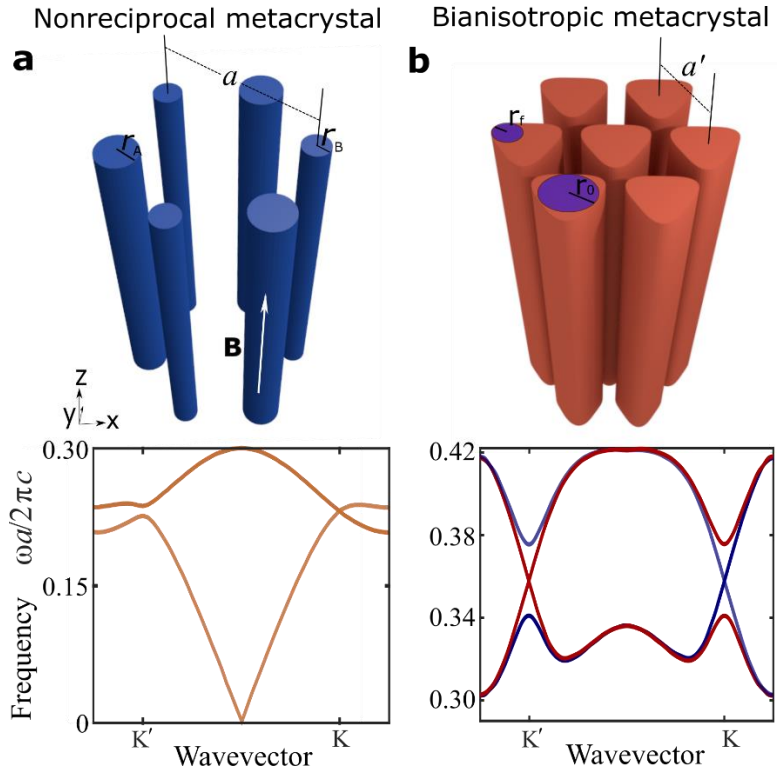

**Figure 1 | Schematic geometries and corresponding band structures** for (a) nonreciprocal 2D photonic crystal composed of ferromagnetic rods arranged in a honeycomb lattice with magnetic bias applied along z direction. (b) Topological 2D photonic crystal made of bi-anisotropic triangulated rods arranged in a triangular array in air. Dimensionless (normalized to lattice constants $a$ and $a'$) geometric parameters of the rods are (a) $r_A = 0.191, \delta = r_A - r_B = 0.01, \mu = 2$, $\kappa = 0.6, \epsilon_A = \epsilon_B = 14$. (b) $r_0 = 0.34, r_f = 0.27, \epsilon_{||} = \mu_{||} = 14, \epsilon_z = \mu_z = 1$. A bianisotropic response $\xi_{xy} = 0.2$ is introduced in the background. In (b), red bands correspond to (pseudo) spin-up states, while blue bands to (pseudo) spin-down states.

## II. One-way Klein Tunneling

Klein tunneling consists in the unimpeded penetration of particles through potential barriers, and it represents one of the most fascinating phenomena associated with the Dirac spectrum. It was first encountered in relativistic quantum physics of high-energy spin-1/2 fermions [49]. More recently, a prediction of this exotic chiral transport found experimental confirmation [50-51] in junctions of graphene [52], which exhibits quasi-relativistic carrier dynamics described by the effective massless Dirac equation [53-55]. Optical analogue of this unique two-dimensional (2D) material, referred to as photonic graphene, also shows unprecedented optical characteristics and it is of eminent research interest in photonics [28].

In the systems under investigation Klein tunneling is expected to assume an even more exotic form – consisting in valley and spin polarized uniform transmission through the Dirac bands. To illustrate such behavior, we performed analytical (Supplement II) and first-principles numerical studies of wave transmission through a photonic potential barrier introduced in the middle of a larger domain. To describe the transmission analytically, we consider a geometry sequence 1/2/1, where domains 1 and 2 are characterized by total masses $m_{1,2}$ and homogeneous potentials $u_{1,2}$. We assume the case of normal incidence $k_y = 0$ of the propagating wave ($|\Omega| > m_1$) from region 1 along the $x$ axis onto the potential barrier of height $(u_2 - u_1)$ and width $L$ located in domain 2, which is assumed to be infinitely long in the $y$ direction. By solving for the wavefunctions in each region with the respective Hamiltonians and then applying the matching (continuity) boundary conditions, we obtain the transmittance

$$T = \frac{4s_1^2|s_2|^2}{|2 s_1 s_2 \cos(k_{2x}L) - i(s_1^2 + s_2^2)\sin(k_{2x}L)|^2}, \qquad (3)$$

where $s_{1,2} = \sqrt{(\Omega - u_{1,2})^2 - m_{1,2}^2} / (\Omega - u_{1,2} - m_{1,2})$, and $k_{2x} = \sqrt{(\Omega - u_2)^2 - m_2^2}/v_D$. Assuming linear dispersion for the K valley, the well appears to be perfectly transparent in K valley for all frequencies, which is a direct manifestation of Klein tunneling. For the K' valley, in contrast, where the spectrum is parabolic, total transmission takes pace only within the passband when the resonant conditions is satisfied, $k_{2x}L = N\pi, N \in \mathbb{Z}$ and, in the gap, $T$ decreases exponentially with the depth (and width) of the well (Supplement II). To verify these analytical predictions, the electromagnetic response of the two proposed crystals in Fig. 1 was modeled using the finite-element method solver in COMSOL Multiphysics.

**One-way valley polarized Klein tunneling in photonic topological crystal with broken TRS**

In Fig. 2(a), top panel, we show the two lowest TM bands of the honeycomb crystal for the case when the magnetic field is absent and the rods in the unit cell are identical. Since time and space inversion symmetries are preserved, the band structure is reciprocal with respect to the wavevector $k_x$ and the band diagram features a pair of Dirac bands with nearly linear dispersion crossing at K $(\frac{4\pi}{3a}, 0)$ and K' $(-\frac{4\pi}{3a}, 0)$ points. In Fig. 2 (a), middle panel, we show the dispersion of the same structure with TRS and SIS broken by applying opposite magnetization in the two cylinders of each unit cell. In this case, the dispersion exhibits two slightly non-reciprocal Dirac cones with slightly detuned frequencies, due to the applied magnetic bias configuration. Importantly, such magnetization preserves parity-time (PT) symmetry, so that Dirac-like dispersion is maintained and the system remains topologically trivial.

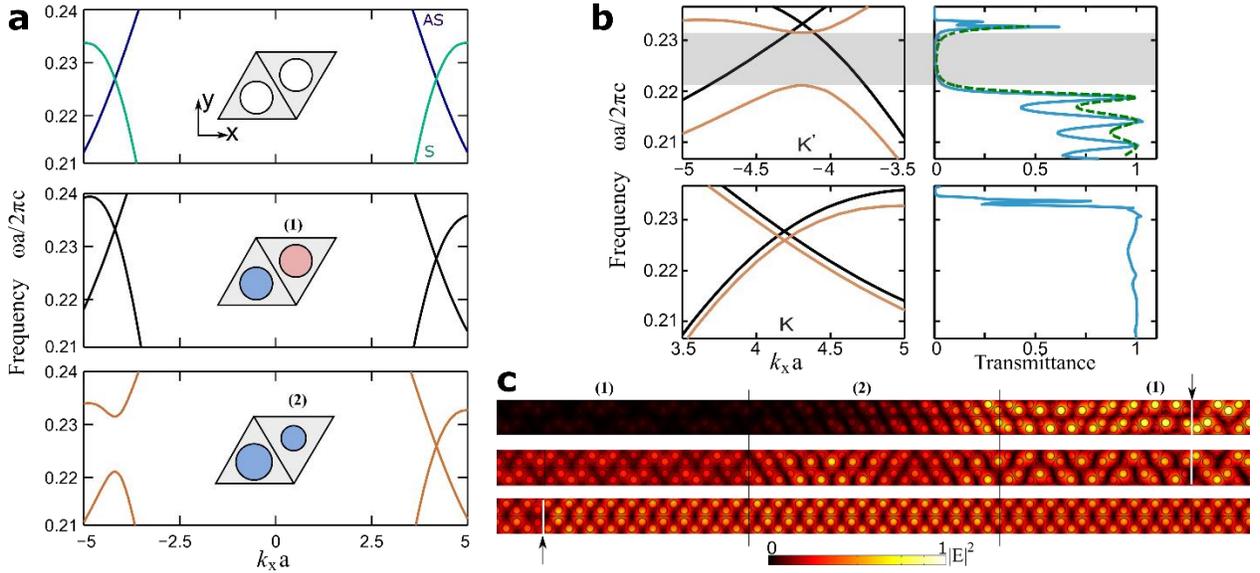

**Figure 2 | One-way Klein tunneling in Nonreciprocal PTI.** (a) Dispersion bands for a symmetric (non-dimerized) structure of non-magnetized rods, $\mu = 2$ (top panel); nonreciprocal PT-preserving crystal, cylinders of equal radii are magnetized in opposite directions (middle panel), $\mu = 2$, $\kappa = 0.6$; nonreciprocal PT-violating crystal, cylinders of slightly detuned radii are magnetized in the same direction (bottom panel), parameters are as in Fig. 1. (b) Photonic bands near K and K' points (left panels) and transmission coefficients (right panels). Top panels and bottom panels correspond to the backward and forward wave propagation, respectively. Numerically calculated transmission is plotted with blue line. The analytically retrieved dependence is shown with a green dashed line. (c) Simulated electric field intensity $|E|^2$ distributions in the strip for backward (top and middle panels) and forward (bottom panel) wave propagation. The strip consists of three domains, domain (1) is the nonreciprocal PT-preserving honeycomb lattices and separated by domain (2) composed of the inequivalent-sites lattice with magnetic field applied perpendicular to the lattice, and domains (1) and (2) contain 2 × 42 and 2 × 12 unit cells, respectively. The boundaries of three crystal regions are marked by black vertical lines. The modes are excited by current sheets at cuts indicated by the arrows and white lines.

Finally, we introduce valley-chirality coupling in the system (lower panel in Fig.2(a)). First, we dimerize the structure, i.e., a small detuning in the cylinders radii $r_{A(B)} = r \pm \delta/2$, is introduced, and SIS is broken, leading to the valley-Hall state. Second TRS symmetry is broken by magnetizing both A and B sites in the same direction, which drives the systems in quantum Hall-like state, and we observe nonreciprocal Dirac spectra where. Next, we tune the parameter $\delta$ such that the effects of magnetization and dimerization cancel out for positive values of $k_x$ (K-valley), thus closing the gap, but amplify the bandgap for negative values of $k_x$ (K′-valley). This one-way nonreciprocal response is shown in the photonic band structure in Fig.2 (a) bottom panel, and it clearly reveals the presence of a one-way Dirac cone.

For the band structures at K′ valley ($k_x < 0$) of two nonreciprocal crystals, plotted in Fig.2(b) upper left panel, we observe an overlay of the Dirac and gapped bands, while for the ones at K valley ($k_x > 0$), plotted in Fig.2(b) lower left panel, two nearly overlapping Dirac cones appear in the spectrum. We use these non-reciprocal photonic band structures to design a large scale photonic transport analogue of Klein tunneling in a supercell that consists of three adjacent domains separated by zigzag cuts. The two side domains are constructed from the nonreciprocal PT-preserved crystal (gapless spectrum at both K and K′ valleys). The middle domain is made of the nonreciprocal PT-violating crystal exhibiting one-way Dirac cone, and it also effectively behaves as a potential well for propagating waves due to the spectral shift of the photonics bands between the domains (Fig. 2(b), lower panel). To calculate the transmission through the potential well, we use a pair of current sheets located in the side regions, which selectively excite modes with positive and negative group velocities at K and K′ valleys, respectively. As expected for Klein tunneling, the calculated transmission is almost unity for the K-valley (forward propagation) across a wide frequency range, as illustrated by Fig.2(b) lower right panel. Only at higher frequencies, where the band becomes parabolic due to the presence of the higher-order band gap, the transmission starts to fall-off. By contrast, one can see that for the K′ valley (backward propagation), Klein tunneling is not observed and the transmission rapidly drops to zero in the bandgap region, which is consistent with the band structure in Fig.2(b), left upper panel. At the same time, transmission through the parabolic passband at K′ valley exhibits numerous resonances intermitted by regions of low transmission. These first-principle results were subsequently fitted with the use of analytic expression Eq. (3) by assuming $m_1 = u_1 = 0$ (and optimizing other fitting parameters) with the result plotted in Fig. 2(b), green dashed line. Analytical and numerical results are in excellent qualitative agreement confirming the origin of the observed valley-polarized one-way Klein tunneling.

Inspection of the field profiles in Fig.2(c) obtained by first-principles simulations at different frequencies, provides additional details explaining the disparate wave transport behavior at the two valleys. Thus, when the structure is excited from the right edge (K′-valley case) of the domain (as indicated by a black arrow), at the frequency within the bandgap, the excited wave clearly experiences an exponential decay in the middle region (Fig.2(c) upper panel). At the lower frequency, a standing wave pattern is clearly formed in the middle region (Fig.2(c) middle panel), explaining the resonant transmission behavior at certain frequencies. Conversely, when the system is excited from the left (K-valley case), a uniform field distribution is observed over the whole structure (Fig.2(c) lower panel). The wave travels towards the right without decay, featuring the analogue of one-way Klein tunneling in this magneto-optical photonic crystal.

# One-way spin-valley polarized Klein tunneling in TRS invariant photonic topological insulator

Switching to the case of TRS invariant case, the second system in Fig. 1, we calculate the photonic band structure of a triangular lattice of circular rods preserving the duality $\epsilon = \mu$, which exhibits a pair of overlaid Dirac spectra at both K and K′ valleys and is plotted in Fig.3(a), top panel. The inversion symmetry reduction is made by triangulating the rods, which leads to lifting the degeneracy at the Dirac points. Bianisotropic response is also introduced to mix TE and TM modes, effectively creating gauge fields with opposite signs for (pseudo) spin-up and spin-down states. These two mechanisms of bandgap opening are tuned by changing the parameters of triangulation and bianisotropy in such a way that for pseudospin up state, plotted with a red line in Fig. 3(a) lower panel, the bandgap closes at K valley and doubles at the K′ valley, thus exhibiting valley dependent Dirac-like and parabolic dispersions, respectively. For the spin-down state, shown with blue lines, the situation is reversed, and the linear dispersion appears at K′ valley instead. One can regard this scenario as the two copies, essentially time reversal partners, of the one-way Dirac cones described above for the magnetic crystal with broken TRS. Therefore, the TRS invariant system should also exhibit one-way Klein tunneling, with an additional selection rule with respect to the spin degree of freedom, thus exhibiting spin-valley dependent transport.

Turning to first-principles simulations of spin-valley polarized transmission, in Fig. 3(b) left panels we plot alongside the band structure for i) the crystals without either gap-opening mechanisms (black lines) and the optimized spin-valley coupled structure with the Dirac cone for spin-up (down) state at one of the valleys. The Dirac cones largely overlap, with a slight shift in frequency, sufficient to provide a potential barrier, and coexist with the gapped spectrum for the opposite spin. Note that the flip of valley from K to K′ results in the reversed situation for the spin states. Large-scale simulations were carried for the supercell consisting of 3 domains, with the middle domain made of the triangulated rods with the bianisotropic response, and two side domains made of non-triangulated and non-bianisotropic rods. Periodic boundary conditions imposed along the top and bottom boundaries, and matching layers are implemented on the sides. To ensure both spin-polarized and valley-polarized excitation, the field source was constructed from a pair of current sheets located at right (left) side domains, as indicated by arrows and white lines in Fig. 3(c). The symmetric current distribution again ensures that we always excite the mode with positive (negative) group velocity at K (K′) valley. Transmission spectra for left (right) to right (left) excitation calculated for each spin are shown in Fig. 3(b), right panels. For the Dirac-bands, the Klein tunneling is observed and a perfect uniform transmission from the left (right) to right (left) is observed for the pseudo spin-up (spin-down) state, while for the opposite spin, which perceives the gapped spectrum at the particular valley, the transmission is suppressed due to exponential decay. Also, as expected for the parabolic bands, the transmission exhibits oscillatory behavior at frequencies outside the bandgap region.

The electric field distributions corresponding to the cases of forward (K valley) and backward (K′ valley) valley-polarized and spin-polarized excitations are plotted along the supercell strip. When the excitation frequency is chosen inside the bandgap, the spin-up (spin-down) state transmitted from left (right) to the right (left), i.e. at K-valley (K'-valley), we observe uniform field intensity distribution as in Fig.3(c), lower panels. At the same time, for opposite spin polarizations excited at the same valleys within the gapped frequency band, the field undergoes exponential decay as seen in Fig.3(c), top panels. For the frequency chosen within the parabolic dispersion outside the bandgap, the field exhibits a clear standing wave pattern shown in the Fig.3(c), middle panels, which complies with the transmission spectra in Fig.3(b). These numerical results therefore vividly demonstrate the most extreme case of spin-valley polarized transport in the form of one-way spin-polarized Klein tunneling.

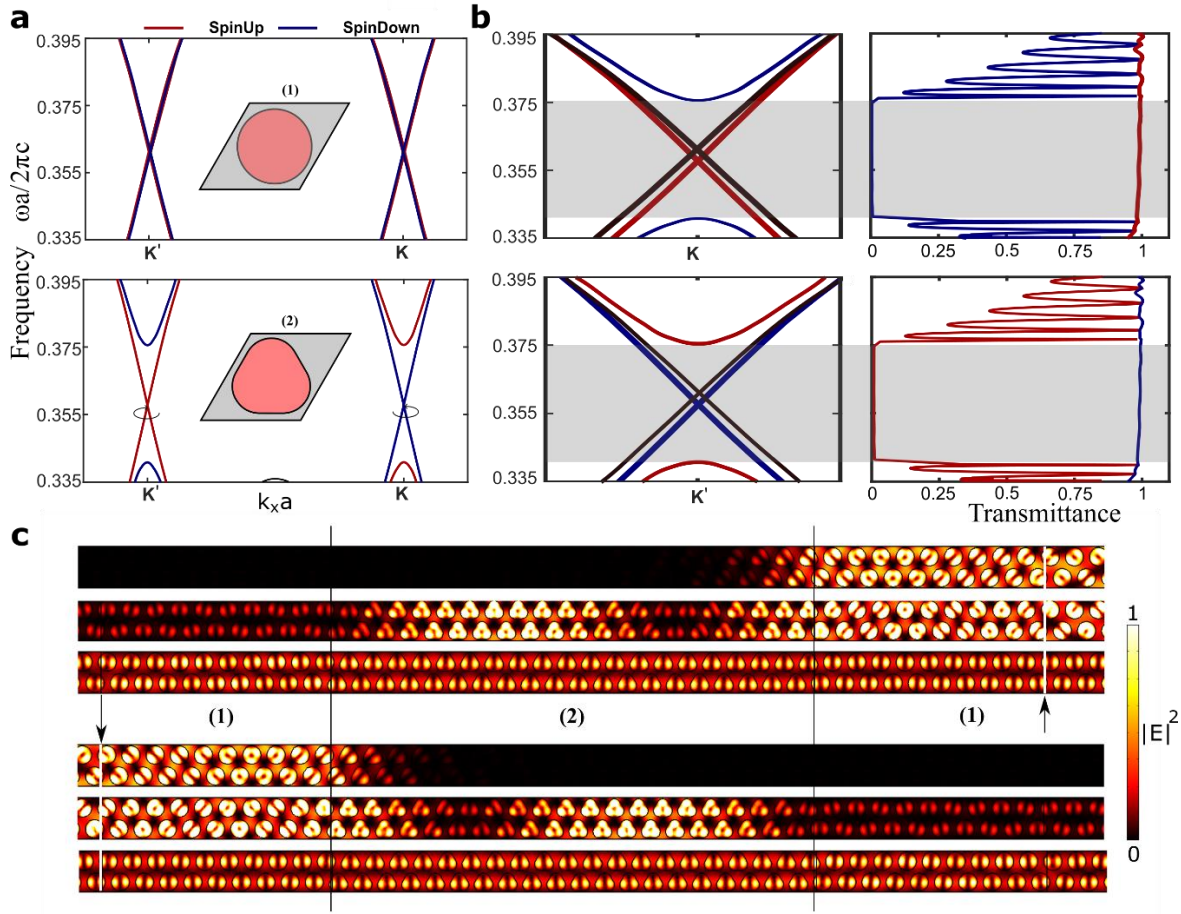

**Figure 3 | Spin-Valley coupled Klein tunneling in bianisotropic PTI.** (a) Photonic band diagrams for dual triangular lattice without any symmetry reduction and with $\epsilon_{\parallel} = \mu_{\parallel} = 14$ (top panel), and triangulated triangular lattice with $\epsilon_{\parallel} = \mu_{\parallel} = 14$ with bianisotropic response $\xi_{xy} = 0.2$ introduced along with fillet of radius $r_f = 0.27$ at each vertex (bottom panel) other parameters are the same as before. Red bands correspond to the states of (pseudo) spin-up, and blue lines to the states of (pseudo) spin-down. (b) Photonic band diagrams of the symmetric triangular lattice (black lines) overlapped with the triangulated rods and bianisotropy introduced. Transmission spectra for the spin-polarized source generating selectively (pseudo) spin-up or spin-down state for the transport at K (K') valleys, as shown in the upper (lower) right panel. (c) Simulated electric field intensity $|\mathbf{E}|^2$ distributions along the strip excited by the sources containing only one pseudo-spin component. The strip consists of three domains, with two domains type (1), which

represent triangular lattices of circular rods with $\epsilon = \mu$, separated by domain type (2), which is composed of the triangulated rods with bianisotropy. The domains (1) and (2) contain $2 \times 25$ and $2 \times 20$ unit cells, respectively. Black dashed lines denote the boundaries between domains. The source is placed in right domain (1) in the upper panel, and in left domain (1) in the lower panel, as indicated by black arrows and white lines.

### III. Edge states in the continuum at critical points of topological transitions

While the observed one-way Klein tunneling appears at the crossover between two topological phases, i) QH-like and VH phases and ii) QSH and VH phases, respectively, it is important to assess whether the structures under investigation can host topological surface states. To answer this question, we performed first-principle numerical studies of the edge states residing in the bandgaps for both TRS breaking and TRS preserving scenarios.

In our simulations for the nonreciprocal crystal, the supercell was chosen to contain a domain wall between two domains possessing mass terms of opposite signs due to simultaneous TRS and SIS reduction, in which case the bandgaps and the Dirac cones for the two domains appeared at the same valleys. We found that the edge mode does occur within the gap at K′ valley, and it coexists spectrally with the gapless bulk continuum at K valley as shown in Fig.4(a). Fig.4(c) shows the results of large-scale numerical modelling, where the edge mode excited by a point dipole source placed at the domain wall clearly propagates one-way at K′-valley in upward direction. At the same time, one can clearly see that the dipole also excites a continuum of bulk modes at K′-valley with the field distributed all over the simulated domains. Note that the crystals were rotated by 90 degrees in these simulations.

Similarly, for the TRS-invariant bianisotropic crystal of triangulated rods, we calculated the band diagram of a supercell made of two domains with mass terms of opposite signs. The latter was achieved by both inverting the orientation of triangulated rods and their bianisotropy, and the band structures found from first principles simulation are shown in Fig. 4(b). Interestingly, edge states supported by the domain wall for the spin-up and spin-down states appear at different valleys, and they coexist with the corresponding bulk continuum of opposite spin state at the same valley. Large-scale simulations enable the observation of such one-way propagation of spin-valley locked edge modes. A source containing both electric and magnetic dipole components was placed at the domain wall to excite only pseudo spin-up or spin-down mode. As seen from Fig. 4(d), the pseudo spin-up (spin-down) edge state excited by the source propagates upwards (downwards) along the domain wall (indicated by the white dashed line), in the direction that corresponds to K-valley (K′-valley), while the spin-up (spin-down) bulk modes propagate in all directions, as expected for the bulk Dirac spectrum.

Simulation results for both TRS breaking and TRS preserving structures therefore unambiguously show that combined action of gauge fields in sublattice and valley, and spin and valley subspaces leads to the appearance of one-way topological edge states that are polarized with respect to both degrees of freedom.

### IV. Proposed experimental design

Both suggested systems in Fig. 1 can be readily implemented in practice. Indeed, topological states and robust transport have been successfully demonstrated experimentally in the recent past for both systems with and without TRS. In particular, a graphene-like photonic crystal with ferrite rods was used to demonstrate edge states confined at the interface with free space [38]. This system would need only a minor modification, the dimerization of ferrite posts, to introduce the effective valley gauge field that can couple to chiral states induced by the magnetization. In order to realize the spin-valley polarized analogue of Klein tunneling, we propose a design that is amenable to a physical implementation in the microwave domain, with details summarized in Supplement III.

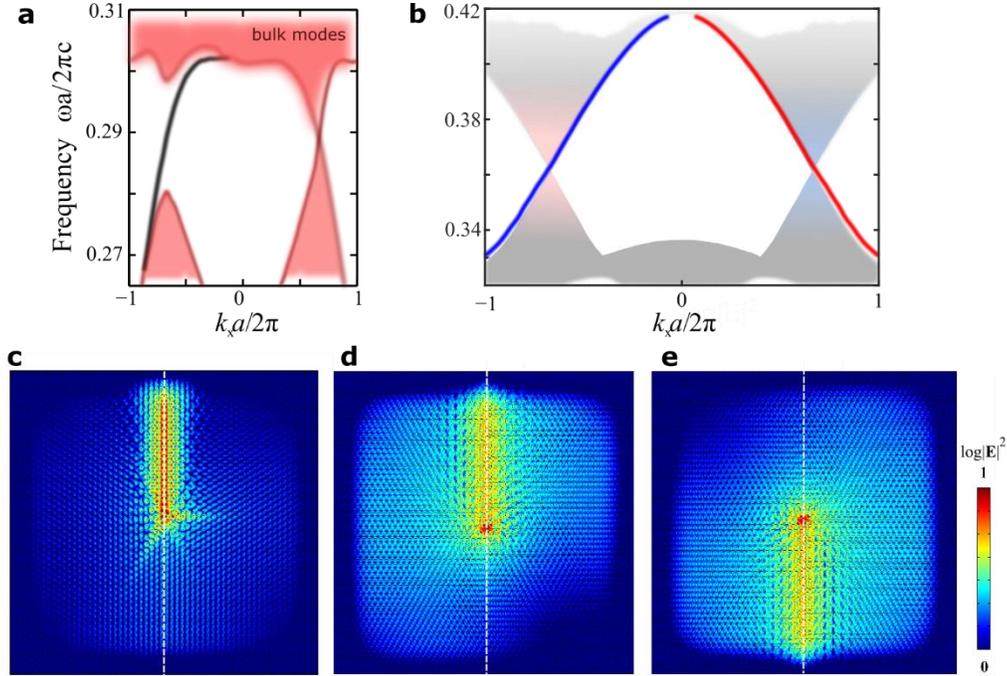

**Figure 4 | Edge states in the continuum**. Bands diagram of a supercell consisting of two domains, which have opposite signs of mass terms for **(a)** TRS reduction and SIS reduction and **(b)** bianisotropy and SIS reduction. **(a)** The continuum of bulk modes is shown by red shaded area and a dispersion branch of the edge mode supported by the domain wall is depicted by black line. **(b)** Grey and color shaded regions show dispersion for continuum of bulk modes. Color shading indicates (pseudo) spin-states of the continuum: blue for spin-down and red for spin-up, respectively. Red and blue solid lines are the bands corresponding to the edge states supported by domain wall for (pseudo) spin-up and spin-down states, respectively. **(c, d, e)** Numerically calculated field distributions in the crystal illustrating excitation of the one-way edge mode at the domain wall. **(c)** The interface separating domains with masses of opposite signs, $m_I = m_T < 0$ (left half-space) and $m_I = m_T > 0$ (right half-space) is marked with a white dashed line. Each domain consists of $30 \times 50$ unit cells. **(d, e)** The region consists of two domains with opposite mass terms, $m_I = m_\chi < 0$ (left half space) and $m_I = m_\chi > 0$ (right half space). Each domain consists of $30 \times 70$ unit cells. The excitation sources consist of current sheets placed in the middle of the domain wall and generating **(d)** spin-up ($E_z + H_z$) and **(e)** spin-down ($E_z - H_z$) components, respectively. To ensure reflectionless propagation of the waves at outer boundaries, we add progressively increasing material losses in the background in the region enclosing the entire simulated structure. Accordingly, darker blue color closer to the boundaries in **(c, d ,e)** implies exponential decay of the fields due to such custom-built crystalline perfectly matching layers.

## V. Conclusion

Synthetic gauge fields acting on either natural or engineered degrees of freedom of light, such as chirality/sublattice, polarization, pseudo-spin and valley, offer an unprecedented degree of control of electromagnetic fields and have already proven to be of great potential in photonics. Combining the effects of such gauge fields, acting in synergy on different synthetic dimensions, allows to significantly expand the space of possible optical responses beyond traditional emulation of topological phases. Here we demonstrated that, in both TRS violating and TRS preserving systems, such combined action of sublattice-valley and spin-valley potentials enables one-way valley and spin-valley polarized transport, respectively. This allows to increase effectively the dimensionality of synthetic photonic potentials thus enabling multi-selective control of electromagnetic radiation. As an example, we demonstrated one-way sublattice-valley and spin-valley polarized Klein tunneling, which represent the extreme case of selective action of such hybrid photonic potentials enabling efficient filtering of bulk modes by either of the degrees of freedom used, as well as a new class of valley polarized one-way and spin-locked edge states. An increasingly large domain of synthetic degrees of freedom engineered in photonics, from quasicrystals [17,21] to synthetic dimensions in multispectral Floquet systems [22,23], makes it even more interesting to investigate the effects of coactive gauge potentials in such complex higher-dimensional systems. To the best of our knowledge, the classification of topological orders in systems with several gauge fields acting on orthogonal internal degrees of freedom is not well understood and it may require the use of non-Abelian gauge field theory methods [23], which makes it of special fundamental interest.

**Methods**:

*Numerical modeling:* The full-wave numerical modeling of electrodynamic response of photonic structures was performed with a FEM solver COMSOL Multiphysics, Radio Frequency module (version 5.2a).

**Acknowledgements**

The work was supported by the National Science Foundation (grants CMMI-1537294 and EFRI-1641069).

**Author contributions**

All authors contributed extensively to the work presented in this paper.

**Competing financial interests**

The authors declare no competing financial interests.